\documentclass[reprint,
 amsmath,amssymb,
 aps,
floatfix,
showkeys
]{revtex4-1}

\usepackage[USenglish]{babel}
\usepackage{slashed}
\usepackage{physics}
\usepackage{graphicx}
\usepackage{dcolumn}
\usepackage{bm}
\usepackage[usenames]{color}
\usepackage{xcolor}
\usepackage{float}

\newcommand{\T}{\tilde}

\newcommand{\be}{\begin{equation}}
\newcommand{\ee}{\end{equation}}
\newcommand{\bea}{\begin{eqnarray}}
\newcommand{\eea}{\end{eqnarray}}
\newcommand{\beas}{\begin{eqnarray*}}
\newcommand{\eeas}{\end{eqnarray*}}
\newcommand{\nn}{\nonumber\\}

\begin{document}

\title{Catalysis and inverse electric catalysis in a scalar theory}


\author{M. Loewe$^{1,2,3}$, D. Valenzuela$^{1}$ and R. Zamora$^{4,5}$ }
\affiliation{%
$^1$Instituto de F\'isica, Pontificia Universidad Cat\'olica de Chile, Casilla 306, Santiago 22,Chile.\\
$^2$Centre for Theoretical and Mathematical Physics, and Department of Physics, University of Cape Town, Rondebosch 7700, South Africa.\\
$^3$Centro Cient\'ifico-Tecnol\'ogico de Valpara\'iso CCTVAL, Universidad T\'ecnica Federico Santa Mar\'ia, Casilla 110-V, Valpara\'iso, Chile.\\
$^4$Instituto de Ciencias B\'asicas, Universidad Diego Portales, Casilla 298-V, Santiago, Chile.\\
$^5$Centro de Investigaci\'on y Desarrollo en Ciencias Aeroespaciales (CIDCA), Fuerza A\'erea de Chile, Casilla 8020744, Santiago, Chile.}%


\begin{abstract}
  
In this article we explore the phase diagram associated to the symmetry breaking of a scalar self interacting theory, induced by temperature and the presence of an external electric field. For such purpose, first we obtain the boson propagator, in the presence of a constant external electric field, both in the weak and strong field strength limits. Novel expansions are derived for the effective potential, valid for the whole range of temperature. We found inverse electric catalysis for the weak field region, i.e. a situation where the critical  temperature diminishes as function of the strength of the electric field whereas for a strong field electric catalysis emerges. These behaviors, in the weak and strong intensity sectors, are valid for all possible values of temperature.  

\end{abstract}


\maketitle

\section{Introduction}\label{sec1}
The analysis of the phase space of Quantum Chromodynamics (QCD) is a central issue in the discussion of hadron physics and strong interactions. It allows us to explore the way in which  deconfinement and chiral symmetry restoration are realized in nature. Other type of phases also occur as, for example, a quarkyonic phase or phases associated to the isospin chemical potential to which we will not refer here.

\noindent
There are two main physical scenarios where these transitions might occur: relativistic heavy ion collisions and inside compact objects like neutron stars. In the first case, the temperature and the strength of the huge magnetic field produced in peripheral collisions are the relevant parameters that determine, for example, the evolution of the critical end point (CEP). Future facilities like NICA and FAIR will probably extend this analysis to regions where finite density effects, expressed through a baryonic chemical potential, start to be relevant. Density and magnetic effects dominate, on the other side, the discussion of neutron stars, being temperature in this case a subdominant effect that can be neglected. Consider \cite{general reviews} for some actual general reviews.

\noindent The analysis of these phase transitions has been carried out by means of several  approaches, as  effective lagrangians, \cite{effective lagrangians}, for example the linear and non-linear sigma models, chiral lagrangians, including also the Polyakov loop as an order parameter, a variety of Nambu-Jona-Lasinio models, that might or not include vector fields, etc \cite{bali01,iranianos,zamora1,zamora2,zamora3,simonov03,aguirre02,tetsuya,dudal04,kevin,gubler,noronha01,morita,Ayala1,morita02,sarkar03,band,nosso1,nosso03,Ayala2,zamora4,zamora5,peng,cohen,tavares}. 

\noindent 
The idea of analyzing the role of  external electric fields as a new effects in the discussion of phase transitions has been recently introduced in literature. Those electric fields appear in asymmetric peripheral relativistic heavy ion collisions, for example Cu + Au collisions \cite{25,26,27}, where in the  overlapping region a strong electric field appears. This is due to the different number of electric charges in each nuclei, inducing the formation of a sort of an electric dipole field during the initial stages of the collision. It is, therefore, natural and relevant to consider these external electric fields in the discussion of phase transitions.
\noindent
Here we will consider, as a toy model, the symmetry breaking pattern of a self interacting theory of charged bosons, a $\lambda \phi ^4$ theory, under the influence of an electric field and temperature.

\noindent
As it is well known, such a discussion proceeds through the determination of the effective potential. The crucial tool for our analysis is the bosonic propagator in the presence of an external electric field. We will present this propagator in the strong and weak field regimes, where strong and weak refer to a comparison of the field strength to the mass. Using these explicit expression, we will calculate our effective potential in both regimes. During this part of the discussion we will present novel expressions for the effective potential valid for the whole range of temperature values. The main conclusion will be the appearance of inverse electric catalysis in the weak field sector whereas in the strong field sector electric catalysis appears. In both cases we found explicit analytic expression valid for all values of temperature. 

\noindent
This article is organized as follows. In section \ref{secII} we derive the bosonic propagator for both  weak and strong electric field intensities.  In section \ref{sec3}  we make use of these propagators finding the critical temperature, where  symmetry is restored, as function of the electric field strength. In Section \ref{sec4} we present our conclusions. Some technical details can be found in in two appendices.
\section{The boson propagator in the weak and strong electric field regimes}\label{secII}

\noindent We start with the proper time representation for a charged boson propagator in the presence of an external electromagnetic field encoded in the $F^{\mu \nu}$ tensor \cite{Dittrich,ahmad} which is given by

\begin{equation}
D(P, F) = \int_{0}^\infty d s e^{-m^2 s}\frac{e^{-s P_{\mu} \left(\frac{\tan Z}{Z}\right)^{\mu \nu}P_{\nu}}}{\sqrt{\det[\cos Z]}}, \label{DPF}
\end{equation}
\noindent where $Z^{\mu \nu} =q F^{\mu \nu}s$, with $q$ the electric charge. Our expressions correspond to the euclidean formulation. If we restrict ourselves to the case where the external field is only given by an electric field pointing along the z axis, we obtain

\begin{equation}
    F ^{\mu \nu}=  E f ^{\mu \nu},
\end{equation}

\noindent where

\begin{equation}
f^{\mu \nu} =
\begin{bmatrix}
0&0&0&-1\\
0&0&0&0\\
0&0&0&0\\
1&0&0&0
\end{bmatrix}.
\end{equation}

\noindent It is easy to verify the following identities

\begin{equation}  
(F ^{2n})^{\mu \nu} =(-1)^n\left(E\right)^{2n}\delta _{\parallel}^{\mu \nu},
\end{equation}

\noindent and

\begin{equation}
(F ^{2n-1})^{\mu \nu }= (-1)^{n+1}\left(E\right)^{2n-1}f^{\mu \nu},
\end{equation}

\noindent where 
\begin{equation}
    \delta _{\parallel} = \begin{bmatrix}
1&0&0&0\\
0&0&0&0\\
0&0&0&0\\
0&0&0&1
\end{bmatrix}.
\end{equation}

\noindent Using the usual series expansion, we find
\begin{equation} 
[\cos Z]^{\mu \nu} = \delta _{\perp}^{\mu \nu}+ \delta _{\parallel}^{\mu \nu}\cos(qE is),
\end {equation}

\noindent where 

\begin{equation}
\delta _{\perp} = \begin{bmatrix}
0&0&0&0\\
0&1&0&0\\
0&0&1&0\\
0&0&0&0
\end{bmatrix},
\end{equation}

\noindent
getting then
\begin{equation}
    \sqrt{ \text{det}[\cos(Z)]} = \vert \cos(qEis)\vert = \vert \cosh(qEs)\vert. \label{cosz}
\end{equation}

\noindent We proceed in the same way with the expansion of $\tan(Z)/Z$ finding

\begin{eqnarray}
\left[\frac{\tan Z}{Z}\right]^{\mu \nu}&=& \frac{\tan (qEis)}{qEis}\delta _{\parallel}^{\mu \nu}
+ \delta _{\perp}^{\mu \nu} \nonumber \\ 
&=& \frac{\tanh (qEs)}{qEs}\delta _{\parallel}^{\mu \nu}
+ \delta _{\perp}^{\mu \nu},\label{tanz}
\end{eqnarray}
Substituting Eq.(\ref{cosz}) and Eq.(\ref{tanz}) in Eq.(\ref{DPF}), it is straightforward to obtain

\begin{equation}D(p) =  \int _{0} ^{\infty}{ds}\frac{e^ {-s\left(\frac{\tanh (qEs)}{qEs} p_{\parallel}^2 +p_{\perp}^2+ m^2 \right)}}{\cosh (qEs)}, \label{propE}
\end{equation} 

where $p_{\parallel}$ and $p_{\perp}$
refer to $(p_{4},0, 0, p_{3})$ and $(0, p_{1},p_{2},0)$, respectively. Note that in the euclidean version $p^{2} = p_{\parallel}^2 + p_{\perp}^2$=$p_4^2+p_3^2+p_1^2+p_2^2$.




\subsection{Weak field}
In order to find the propagator in the weak electric field approximation, we will follow the ideas presented in   \cite{ayalaeuro}. A Taylor's series expansion of Eq.(\ref{propE}) for $qEs \sim 0$, gives us
\begin{eqnarray}
&&D(p)\approx\int _{0} ^{\infty}{ds}e^ {-s(p^2+ m^2)}\Biggl[ \frac{(qEs)^2}{6}\left(-3+2sp_{\parallel}^2\right) \nonumber \\
&&+\frac{(qEs)^4 }{360}\biggl(4sp_{\parallel}^2[5sp_{\parallel}^2-27]+75\biggr) +\frac{(qEs)^6}{45360} \nonumber \\
&\times&\biggl(2sp_{\parallel}^2[14 p_{\parallel}^2s(10p_{\parallel}^2s-117)+4311]-3843\biggr)\dots \Biggr], \nonumber \\
\end{eqnarray}
proceeding then to integrate in the proper time parameter $s$. It can be shown  that the propagator, written here up to $\mathcal{O}(E^6)$, becomes 
\begin{eqnarray}
&&D(p)\approx \frac{1}{p^2+m^2} \nonumber \\
&&+(qE)^2\left(-\frac{1}{(p^2+m^2)^3}+\frac{2  p_{\parallel}^2}{(p^2+m^2)^4}\right) \nonumber \\
&+& (qE)^4\left( \frac{5}{(p^2+m^2)^5}-\frac{36  p_{\parallel}^2}{(p^2+m^2)^6}+\frac{40  p_{\parallel}^4}{(p^2+m^2)^7}\right) \nonumber \\
&+& (qE)^6\Biggl(- \frac{61}{(p^2+m^2)^7}+\frac{958  p_{\parallel}^2}{(p^2+m^2)^8}-\frac{2912  p_{\parallel}^4}{(p^2+m^2)^9} \nonumber \\
&+&\frac{2240 p_{\parallel}^6}{(p^2+m^2)^{10}} \Biggr). \label{debil6}
\end{eqnarray}
We notice that a comparison with the weak field expansion for the propagator in the presence of a magnetic field \cite{taiwaneses} shows that $p_{\parallel}^2\leftrightarrow p_{\perp}^2$.
We might understand this in a quite natural way since magnetic phenomena take place in a plane perpendicular to the magnetic field direction whereas the electric dynamics is longitudinal, i.e parallel to the direction of the field. 


    

\subsection{Strong field}
In order to calculate the strong field case, we first use the following identities 
\begin{eqnarray}
\cosh{(x)}&=&\frac{1+e^{-2x}}{2 e^{-x}}=\frac{1+u}{2u^{1/2}} \nonumber \\
\tanh{(x)}&=&1-\frac{2e^{-2x}}{1+e^{-2x}}=1-\frac{2u}{1+u}, \label{trigo}
\end{eqnarray}
where $u=e^{-2x}$.Replacing Eq.(\ref{trigo}) in Eq.(\ref{propE}), we obtain
\begin{eqnarray}
D(p) &=& 2 \int _{0} ^{\infty}{ds}e^{-s( p_{\perp}^2+ m^2)} e^{-\frac{p_{\parallel}^2}{qE}} \nonumber \\
&\times& \frac{u^{1/2}}{1+u}e^{\frac{2p_{\parallel}^2}{qE} \left(\frac{u}{1+u}\right)}, 
\end{eqnarray}
 where we recognize the appearance of the generating function for the Laguerre polynomials \cite{abramowitz}, given by
\begin{equation}
    \frac{e^{-xz/(1-z)}}{1-z}=\sum_0^{\infty}L_l(x) z^l.
\end{equation}
Therefore, we get
\begin{eqnarray}
D(p) &=& 2 \sum_0^{\infty}(-1)^l L_l\left(\frac{2p_{\parallel}^2}{qE} \right)e^{-\frac{p_{\parallel}^2}{qE}} \nonumber \\
&\times&\int _{0} ^{\infty}{ds}e^{-s( p_{\perp}^2+ m^2+(2l+1)qE)}. 
\end{eqnarray}
Finally, we integrate in $s$ obtaining

\begin{equation} \label{LLL}
    D(p)= 2 \sum_{l=0}^{\infty}(-1)^l \frac{L_{l}\left(\frac{2p_{\parallel}^2}{qE}\right)e ^{-\frac{ p_{\parallel}^2}{qE}}}{p_{\perp}^2 +(2l+1)qE + m^2}.
\end{equation}
By comparing with the strong magnetic field case \cite{mexicanos}, we realize again the analogy mentioned before, i.e. $p_{\parallel}^2\leftrightarrow p_{\perp}^2$. The strong field case is obtained for $ l = 0 $  obtaining 
\begin{equation}
    D(p) \approx 2 \frac{e^{-\frac{ p_{\parallel}^2}{qE}}}{p_{\perp}^2 +qE + m^2}.
\end{equation}

\section{Effective potential and symmetry restoration}\label{sec3}
In order to explore the consequences of this electric field scenario, we will use the Abelian Higgs model.
The model is given by the Lagrangian 
\bea
   {\mathcal{L}}=(D_{\mu}\phi)^{\dag}D^{\mu}\phi
   +\mu^{2}\phi^{\dag}\phi-\frac{\lambda}{4}   
   (\phi^{\dag}\phi)^{2},
\label{lagrangian}
\eea
where $\phi$ is a charged scalar field and
 \bea
   D_{\mu}=\partial_{\mu}+iqA_{\mu},
\label{dcovariant}
\eea
We choose $A_{\mu}=-\delta_{\mu 0} x_3 E$ in order to obtain a constant electric field in the z-direction, being $q$ the particle's electric charge. The squared mass parameter $\mu^2$ and the self-coupling $\lambda$ are taken to be positive.

We can write the complex field $\phi$ in terms of the real components $\sigma$ and $\chi$,
\bea
   \phi(x)&=&\frac{1}{\sqrt{2}}[\sigma(x)+i \chi(x)],  \nonumber \\
   \phi^{\dag}(x)&=&\frac{1}{\sqrt{2}}[\sigma(x)-i\chi(x)].
\label{complexfield}
\eea
Following the usual procedure to allow for a spontaneous symmetry breaking, the $\sigma$ field develops a vacuum expectation value $v$
\bea
   \sigma \rightarrow \sigma + v,
\label{shift}
\eea
which can later be taken as the order parameter of the theory. After this shift, the Lagrangian can be rewritten as
\bea
   {\mathcal{L}} &=& -\frac{1}{2}[\sigma(\partial_{\mu}+iqA_{\mu})^{2}\sigma]-\frac{1}
   {2}\left(\frac{3\lambda v^{2}}{4}-\mu^{2} \right)\sigma^{2}\nn
   &-&\frac{1}{2}[\chi(\partial_{\mu}+iqA_{\mu})^{2}\chi]-\frac{1}{2}\left(\frac{\lambda v^{2}}{4}-   
   \mu^{2} \right)\chi^{2}+\frac{\mu^{2}}{2}v^{2}\nn
  &-&\frac{\lambda}{16}v^{4}
  +{\mathcal{L}}_{I},
  \label{lagranreal}
\eea
where ${\mathcal{L}}_{I}$ is given by
\bea
  {\mathcal{L}}_{I}&=&-\frac{\lambda}{16}\left(\sigma^4+\chi^4+2\sigma^2\chi^2\right),
  \label{lagranint}
\eea
and represents the interactions among the  $\sigma$ and $\chi$ fields, after symmetry breaking. It is well known that in the Abelian Higgs model the gauge field $A^\mu$ acquires a finite mass and thus does not represent a massless photon interacting with the charged scalar field \cite{Nambu}. Therefore, for our discussion we will ignore the mass term generated for $A^\mu$, as well as issues regarding renormalization after symmetry breaking, and will concentrate on the scalar sector. From Eq.~(\ref{lagranreal}) we see that the $\sigma$ and $\chi$ masses are given by
\bea
  m^{2}_{\sigma}&=&\frac{3}{4}\lambda v^{2}-\mu^{2},\nn
  m^{2}_{\chi}&=&\frac{1}{4}\lambda v^{2}-\mu^{2}.
\label{masses}
\eea
To lowest order (tree level) the potential is
\begin{equation}
  V^{\text{(tree)}}= -\frac{1}{2}\mu^2 v^2 +\frac{1}{16}\lambda^2 v^4,
\end{equation}
 Going now into radiative corrections, the one-loop effective potential for a self interacting scalar field at finite temperature, in the presence of a constant electric field, can be written as \cite{LeBellac,Kapusta}
\begin{eqnarray}
    V^{\text{(\text{1-loop})}}&=&\sum_{i=\sigma,\chi}\Biggl(\frac{T}{2}\sum_n \int \frac{d^3p}{(2\pi)^3} \ln[D(\omega_n,p,m_i)]^{-1}\Biggl) \nonumber\\
    &=&\sum_{i=\sigma,\chi}\Biggl( \frac{T}{2}\sum_n \int dm_i^2 \int \frac{d^3p}{(2\pi)^3} D(\omega_n,p,m_i)\Biggl), \nonumber \\ \label{efectivo}
\end{eqnarray}
where finite temperature effects have been introduced by means of the imaginary time formalism in the usual way, i.e.
\begin{equation}
p_4 \rightarrow \omega_{n} = 2\pi n T,\; n\in Z.
\end{equation} 
 and where the integral in $p_4$ converts into a sum according to,
\begin{eqnarray}
\int \frac{d^4p}{(2\pi)^4}f(p) \rightarrow T\sum_{n\in Z} \int \frac{d^3k}{(2\pi)^3} f(\omega_n,p).
\end{eqnarray}
In order to  calculate the Eq.~(\ref{efectivo}) analytically, we will analyze the weak and strong electric field regimes separately.
\subsection{Weak field regime}

\begin{figure}[h]
\includegraphics[width=85mm]{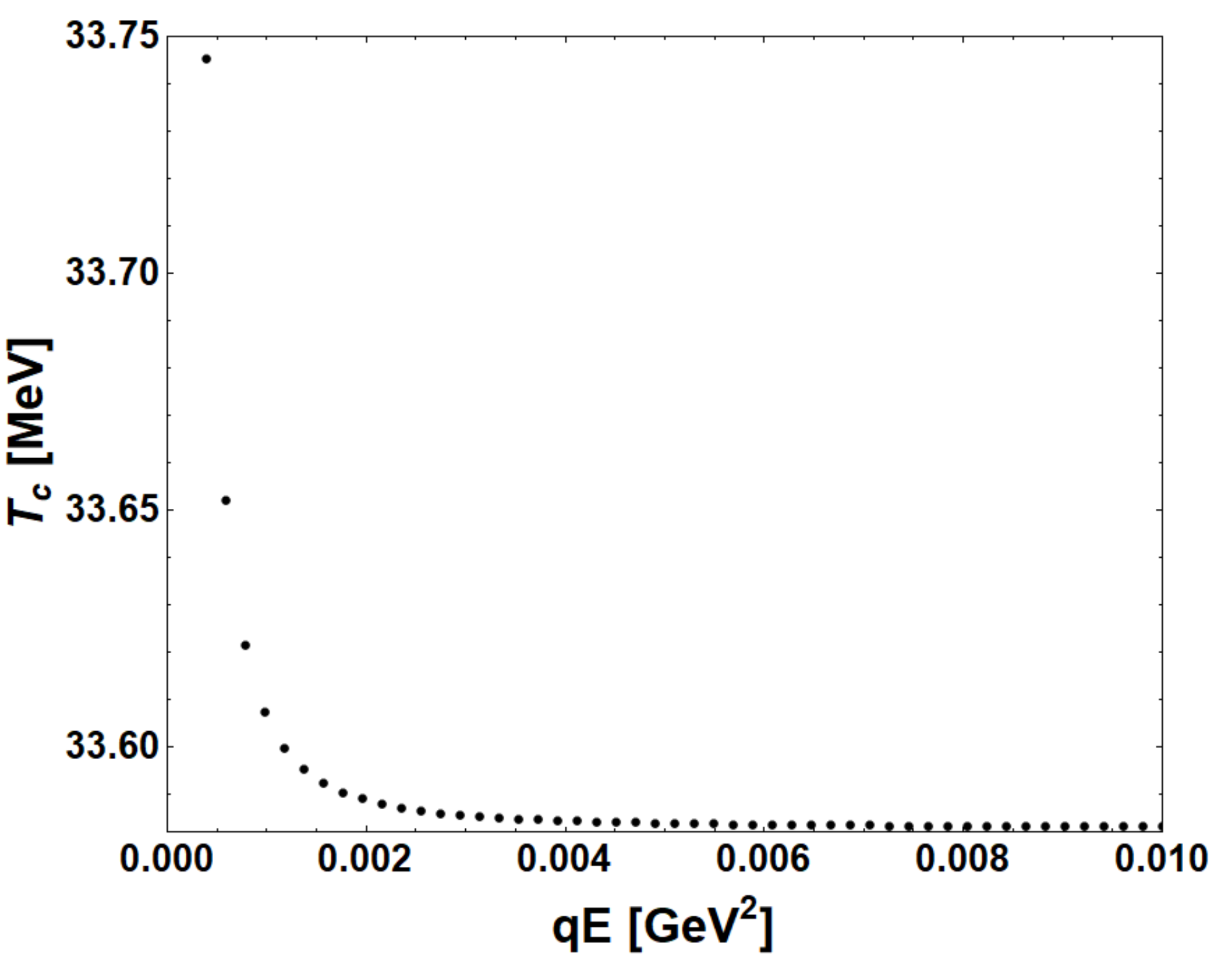} 
\caption{Critical  temperature behavior in the weak electric field region.}
\label{criticadebil}
\end{figure}

We will calculate the effective potential, in the weak field regime, for any value of temperature. In the corresponding discussion for the weak magnetic field case, the usual analysis is done for high temperature, where inverse magnetic catalysis appears. In the present case we will be able to calculate it analytically for the whole range of temperature. For this we will consider the propagator given by   Eq.(\ref{debil6}), only up to order  $\mathcal{O}(E^2)$. Therefore, substituting  Eq.(\ref{debil6}), up to this order, into Eq.(\ref{efectivo}) we get
\begin{eqnarray}
    &&V_{\text{weak}}^{\text{(\text{1-loop})}}=\sum_{i=\sigma,\chi}\Biggl[\frac{T}{2}\sum_n \int dm_i^2 \int \frac{d^3p}{(2\pi)^3} \frac{1}{\omega_n^2+p^2+m_i^2} \nonumber \\
    &+&(qE)^2\left(\frac{2(\omega_n^2+p_3^2) }{(\omega_n^2+p^2+m_i^2)^4} -\frac{1}{(\omega_n^2+p^2+m_i^2)^3}\right)\Biggr]. \nonumber \\
\end{eqnarray}
In appendix A we present the details how to handle the previous expression, for the the whole range of temperature, since there are several subtle steps involved in the calculation which demand also the use of some interesting and novel techniques. We obtain  
\begin{eqnarray}
&&V_{\text{weak}}^{\text{(\text{1-loop})}}=\sum_{i=\sigma,\chi}\Biggl(-\frac{m_i^4}{64\pi^2} \biggl(
\ln\left(\frac{\widetilde{\mu}^2}{m_i^2}\right)+\frac{3}{2} \biggr) \nonumber \\
\nonumber \\
&-&\frac{m_i^2 T^2}{2\pi^2} \sum_{n=1}^\infty \frac{K_2(nm_i/T)}{n^2}+ \frac{(qE)^2}{96 \pi^2}       \Biggl[\ln\left(\frac{\widetilde{\mu}^2}{m_i^2}\right)   \nonumber \\
&&  +1 + \sum_{n=1}^{\infty} 2K_0(nm_i/T) + \frac{2m_i}{T} \sum_{n=1}^{\infty}nK_1(nm_i/T)  \Biggr]\Biggr), \nonumber \\
\end{eqnarray}
where $\widetilde{\mu}$ is the the ultraviolet renormalization scale. Therefore, the effective potential  for the case of a weak electric field, up to the 1-loop order, can be written for any temperature as
\begin{eqnarray}
&&V_{\text{weak}}=V^{\text{(tree)}}+ V_{\text{weak}}^{\text{(1-loop)}} =-\frac{1}{2}\mu^2 v^2 +\frac{1}{16}\lambda^2 v^4 \nonumber \\
&+&\sum_{i=\sigma,\chi}\Biggl(-\frac{m_i^4}{64\pi^2} \biggl(
\ln\left(\frac{\widetilde{\mu}^2}{m_i^2}\right)+\frac{3}{2} \biggr) \nonumber \\
\nonumber \\
&-&\frac{m_i^2 T^2}{2\pi^2} \sum_{n=1}^\infty \frac{K_2(nm_i/T)}{n^2}+ \frac{(qE)^2}{96 \pi^2}       \Biggl[\ln\left(\frac{\widetilde{\mu}^2}{m_i^2}\right)   \nonumber \\
&&  +1 + \sum_{n=1}^{\infty} 2K_0(nm_i/T) + \frac{2m_i}{T} \sum_{n=1}^{\infty}nK_1(nm_i/T)  \Biggr] \Biggr). \nonumber \\
\end{eqnarray}
In order to find the critical temperature above which the symmetry is restored, the idea is to explore the evolution with temperature of the effective potential as function of the vacuum expectation value ($v$). The effective potential becomes more flat when temperature rises. At the critical temperature the first and second derivatives, in fact all derivatives, vanish becoming then  convex, with a parabolic shape,  for higher temperatures. This is what we actually found. The phase transition is of second order since no degenerated vacua appear in this procedure. Therefore, we can get the behavior of the critical temperature as function of the intensity of the electric field, in the weak region as it is shown in Fig.~\ref{criticadebil}.

\subsection{Strong field regime}
\begin{figure}[h]
\includegraphics[width=85mm]{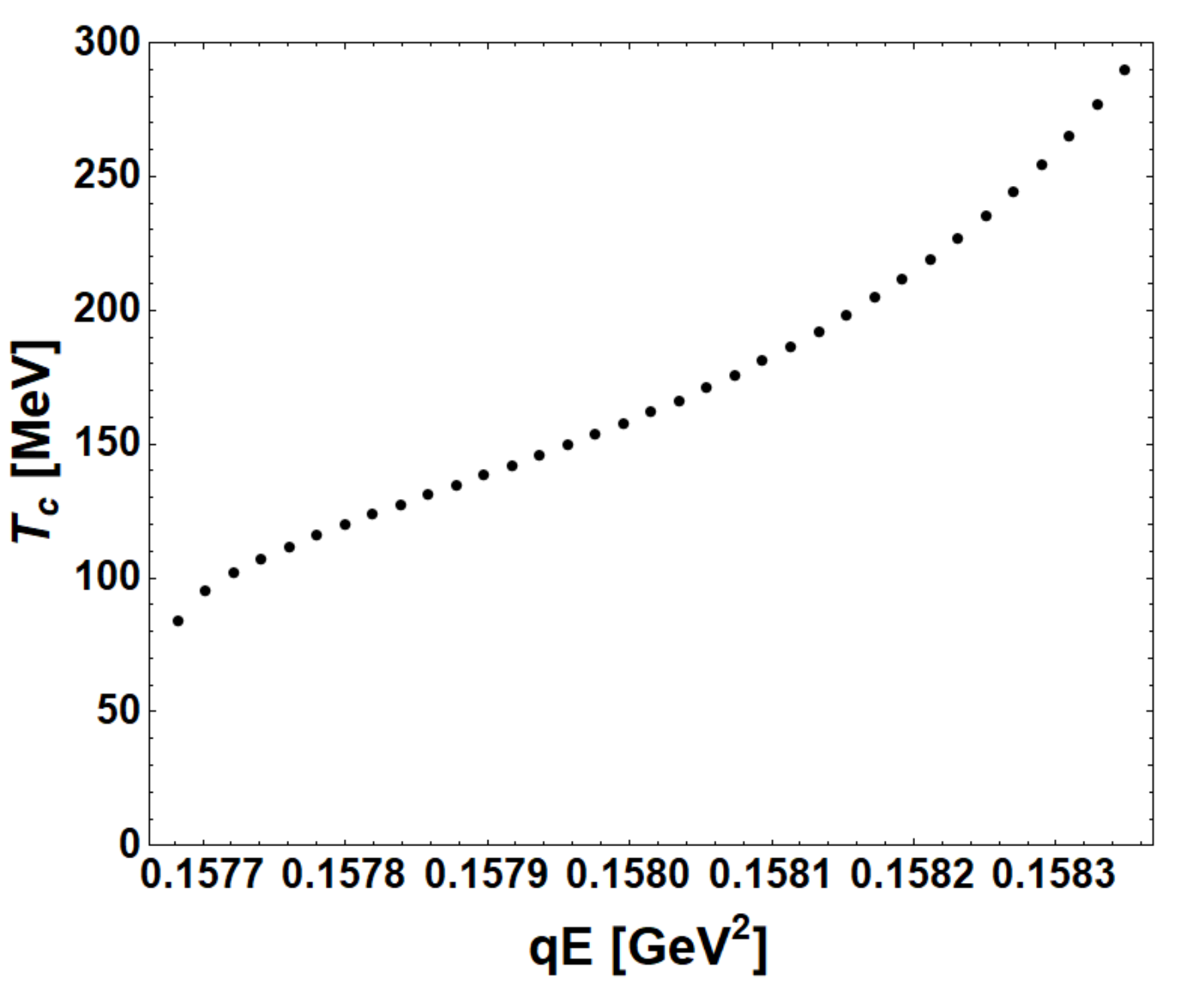} 
\caption{Critical temperature behavior  in the strong electric field region.}
\label{criticafuerte}
\end{figure}

To calculate the effective potential in the strong electric field region, we use Eq.~(\ref{LLL}) in Eq.~(\ref{efectivo}) obtaining
\begin{eqnarray}
V_{\text{strong}}^{\text{(\text{1-loop})}} &=& \sum_{i=\sigma,\chi}\Biggl(\frac{T}{2}\sum_n \int dm_i^2 \nonumber \\
&\times&\int \frac{d^3p}{(2\pi)^3} 2 \frac{e^{-\frac{ \omega_n^2+p_3^2}{qE}}}{p_{\perp}^2 +qE + m_i^2}\Biggr)\nonumber \\
&=&\sum_{i=\sigma,\chi}\Biggl(T \sum_n e^{-\frac{(2\pi n T)^2}{qE}} \int dm_i^2  \nonumber \\
&\times& \int\frac{d^3p}{(2\pi)^3} \frac{e^{-\frac{p_3^2}{qE}}}{p_1^2+p_2^2 +qE + m_i^2}\Biggr) \nonumber \\
&=&\sum_{i=\sigma,\chi}\Biggl(T\vartheta_3\left(0,e^{-\frac{4\pi^2T^2}{qE}}\right) \int dm_i^2  \nonumber \\ 
&\times&\int\frac{d^3p}{(2\pi)^3} \frac{e^{-\frac{p_3^2}{qE}}}{p_1^2+p_2^2 +qE + m_i^2}\Biggr),\nonumber \\
\end{eqnarray}
where we have used Jacobi´s theta function \cite{abramowitz}, defined by
\begin{equation}
    \vartheta_3(z,q)=\sum_{n=-\infty}^{\infty}q^{n^2}e^{2niz}.
\end{equation}
Now we integrate in  $dp_3$, $dp_2$ and $dm^2_i$ obtaining
\begin{eqnarray}
V_{\text{strong}}^{\text{(\text{1-loop})}}&=&\sum_{i=\sigma,\chi}\Biggl(\T\vartheta_3\left(0,e^{-\frac{4\pi^2T^2}{qE}}\right) \sqrt{\frac{qE}{\pi}} \nonumber \\
&\times& \int \frac{dp_1}{2\pi}\sqrt{p_1^2+qE+m_i^2}\Biggr).  
\end{eqnarray}
For the integral in  $dp_1$ we use dimensional regularization,  introducing the ultraviolet renormalization scale $\widetilde{\mu}$ in the $\overline{\text{MS}}$ scheme  and the inclusion of a mass counterterm $\delta m^2\sim m^2/\epsilon$. In this way we obtain
\begin{eqnarray}
V_{\text{strong}}^{\text{(\text{1-loop})}}&=& \sum_{i=\sigma,\chi}\Biggl[ \frac{T\sqrt{\pi qE}}{4\pi^2} \vartheta_3\left(0,e^{-\frac{4\pi^2T^2}{qE}}\right)(m_i^2+qE) \nonumber\\
&\times&\Biggl(1+\ln\left(\frac{\tilde{\mu}^2}{m_i^2+qE}\right)\Biggr)\Biggr].
\end{eqnarray}
Therefore, the effective potential, in the case of a strong electric field, can be written for any value of temperature as 
\begin{eqnarray}
V_{\text{strong}}&=&V^{\text{(tree)}}+ V_{\text{strong}}^{\text{(1-loop)}} \nonumber \\
 &=&-\frac{1}{2}\mu^2 v^2 +\frac{1}{16}\lambda^2 v^4 \nonumber \\
&+&\sum_{i=\sigma,\chi}\frac{T\sqrt{\pi qE}}{4\pi^2} \vartheta_3\left(0,e^{-\frac{4\pi^2T^2}{qE}}\right)(m_i^2+qE) \nonumber\\
&\times&\Biggl(1+\ln\left(\frac{\tilde{\mu}^2}{m_i^2+qE}\right)\Biggr).
\end{eqnarray}
With this strong field effective potential we obtain, as we already did in the case of the weak electric field sector, the evolution of the critical temperature as function of the electric field strength, this time for high field intensities. Our result is shown in  Fig.~\ref{criticafuerte}.

\section{Conclusions}\label{sec4}

From  Fig.~\ref{criticadebil}  we see that we have obtained inverse electric catalysis in the weak electric field region since, when turning the electric field on, the critical temperature starts to diminish. However, the rate of this diminishing  becomes smaller when the electric field starts to grow, showing a damped behavior. This behavior is analogous to what happens in the pure magnetic case \cite{inverse1,inverse2,inverse3,inverse4,inverse5,inverse6,inverse7}. However,when going into the strong electric field regime,  Fig.~\ref{criticafuerte}, we notice that the evolution is just the opposite, appearing electric catalysis. The critical temperature raises as function of the strength of the electric field, being this growing behavior much steeper than the diminishing rate in the weak field sector. These behaviors coincide with the results reported  in \cite{Farias}. In this article the authors used a two-flavor Nambu-Jona-Lasinio model as a theoretical framework. 

It would be interesting to discuss  the case where an electric and a magnetic field are simultaneously present. Regarding the evolution of the critical temperature, probably there will a competition between these two effects. We will report on this situation in a near future.

\section*{Acknowledgements}
M. Loewe and R. Zamora acknowledge support from   ANID/CONICYT FONDECYT Regular (Chile) under Grant No. 1200483. M.L.  acknowledges support from FONDECYT Regular under grant No. 1190192. ML acknowledges also support from ANID PIA/APOYO AFB 180002 (Chile).


\begin{appendix}
\section{Effective potential in the presence of an external electric field for all temperature ranges}\label{apendice}
The calculation will be done for a generic boson of mass $m$. The effective potential up to order $\mathcal{O}(E^2)$ in the regime of a weak electric field at finite temperature is given by
\begin{eqnarray}
    &&V_{\text{weak}}^{\text{(\text{1-loop})}}=\frac{T}{2}\sum_n \int dm^2 \int_{-\infty}^{\infty} \frac{d^3p}{(2\pi)^3} \frac{1}{\omega_n^2+p^2+m^2} \nonumber \\
    &+&(qE)^2\left(\frac{2(\omega_n^2+p_3^2) }{(\omega_n^2+p^2+m^2)^4} -\frac{1}{(\omega_n^2+p^2+m^2)^3}\right)\nonumber \\
    &\equiv& V_{I}+V_{II},
    \end{eqnarray}
    where
    \begin{equation}
    V_{I}\equiv\frac{T}{2}\sum_n \int dm^2 \int_{-\infty}^{\infty} \frac{d^3p}{(2\pi)^3} \frac{1}{\omega_n^2+p^2+m^2}   ,  
    \end{equation}and
    \begin{eqnarray}
    V_{II} &\equiv& \frac{T}{2}\sum_n \int dm^2 \int_{-\infty}^{\infty} \frac{d^3p}{(2\pi)^3}(qE)^2 \nonumber \\ 
    &\times&\left(\frac{2(\omega_n^2+p_3^2) }{(\omega_n^2+p^2+m^2)^4} -\frac{1}{(\omega_n^2+p^2+m^2)^3}\right).   
    \end{eqnarray}
    First we proceed with the calculation of the term  $V_I$.For this, first we carry out the sum of Matsubara frequencies \cite{LeBellac,Kapusta}
\begin{eqnarray}
T\sum_n \frac{1}{(\omega_n^2+p^2+m^2)}&=& \frac{1}{2 \sqrt{p^2+m^2}} \nonumber \\
&\times&(1+2n_B(\sqrt{p^2+m^2})),
\end{eqnarray}
where $n_B$ is the  Bose-Einstein distribution
\begin{equation}
 n_B(x)=\frac{1}{e^{x/T}-1}.   
\end{equation}
Thus, we obtain
\begin{eqnarray}
   V_I&=&\frac{1}{4}\int dm^2\int_{-\infty}^{\infty} \frac{d^3p}{(2\pi)^3}\frac{1}{\sqrt{p^2+m^2}} \nonumber \\
   &\times&(1+2n_B(\sqrt{p^2+m^2})).\nonumber \\
   &\equiv& V_I^{\text{vacuum}}+V_I^{T},\label{vi}
\end{eqnarray}
where
\begin{eqnarray}
&&V_I^{\text{vacuum}}=\frac{1}{4}\int dm^2\int_{-\infty}^{\infty} \frac{d^3p}{(2\pi)^3}\frac{1}{\sqrt{p^2+m^2}} \hspace{1cm} \text{and} \nonumber \\
&&V_I^{T}=\frac{1}{4}\int dm^2\int_{-\infty}^{\infty} \frac{d^3p}{(2\pi)^3}\frac{1}{\sqrt{p^2+m^2}} 2n_B(\sqrt{p^2+m^2}), \nonumber \\ \label{A7}
\end{eqnarray}
The term $V_I^{\text{vacuum}}$ corresponds to the vacuum contribution which is handled through dimensional regularization. Therefore, after taking the ultraviolet renormalization scale $\widetilde{\mu}$ in the $\overline{\text{MS}}$ scheme and the inclusion of a mass counterterm $\delta m^2\sim m^2/\epsilon$, we obtain
\begin{eqnarray}
  V_I^{\text{vacuum}}=-\frac{m^4}{64\pi^2} \biggl(
\ln\left(\frac{\widetilde{\mu}^2}{m^2}\right)+\frac{3}{2} \biggr).
\end{eqnarray}
The term $V_I^{T}$ is a pure thermal contribution
\begin{eqnarray}
  V_I^{T}   &=&\int dm^2 \int_{0}^{\infty} \frac{dp~p^2}{(2\pi)^2} \frac{ n_B(\sqrt{p^2+m^2})}{\sqrt{p^2+m^2}} \nonumber \\
  &=&\int dm^2 \sum_{n=0}^{\infty} \int_{0}^{\infty} \frac{dp~p^2}{(2\pi)^2} \frac{e^{-(n+1)\sqrt{p^2+m^2}/T}}{\sqrt{p^2+m^2}}.
 \end{eqnarray}
 Introducing the change of variable  $p=m \sinh(u)$ we get
 \begin{eqnarray}
    V_{I}^{T}&=&\int dm^2 \sum_{n=0}^{\infty} \int_{0}^{\infty} \frac{du m^2\sinh(u)^2}{(2\pi)^2} e^{-(n+1)m \cosh(u)/T}. \nonumber \\
     \end{eqnarray}
     Using $\cosh(u)^2-\sinh(u)^2=1$, we have
      \begin{eqnarray}
    &&V_{I}^{T}=\int dm^2 \sum_{n=0}^{\infty} \int_{0}^{\infty} \frac{du m^2(\cosh(u)^2-1)}{(2\pi)^2} e^{-(n+1)m \cosh(u)/T} \nonumber \\
    &=&\int dm^2 \frac{m^2}{(2\pi)^2}\sum_{n=0}^{\infty} \Biggl[ \frac{\partial^2}{\partial m^2} \left(\frac{T}{n+1}\right)^2 \int_{0}^{\infty} du e^{-(n+1)m \cosh(u)/T}  \nonumber \\
    &-&\int_{0}^{\infty} du e^{-(n+1)m \cosh(u)/T} \Biggr]. \label{vit}
    \end{eqnarray}
We notice at this point that the modified Bessel function $K_{\alpha}(x)$ has the form \cite{abramowitz}
\begin{equation}
 K_{\alpha}(x)=\int_0^{\infty}du e^{-x \cosh(u)}\cosh(\alpha u).\label{bessel}   \end{equation}
Thus we can write Eq. (\ref{vit}) as 
  \begin{eqnarray}
    V_{I}^{T}&=&\int dm^2 \frac{m^2}{(2\pi)^2}\sum_{n=0}^{\infty} \Biggl( \frac{\partial^2}{\partial m^2} \left(\frac{T}{n+1}\right)^2 K_0((n+1)m/T)  \nonumber \\
    &-&K_0((n+1)m/T)  \Biggr)\nonumber \\
    &=&\int dm^2 \frac{m^2}{(2\pi)^2}\sum_{n=0}^{\infty} \Biggl(\frac{1}{2} K_0((n+1)m/T) +K_2((n+1)m/T)  \nonumber \\
    &-&K_0((n+1)m/T)  \Biggr) \nonumber \\
      &=&\int dm^2 \frac{m^2}{(2\pi)^2}\sum_{n=0}^{\infty} \frac{T}{m(n+1)}   K_1((n+1)m/T) \nonumber \\
       &=&\frac{1}{(2\pi)^2}\int dm^2 m T \sum_{n=1}^{\infty} \frac{K_1(nm/T)}{n}\nonumber \\
&=&-\frac{m^2 T^2}{2\pi^2} \sum_{n=1}^\infty \frac{K_2(nm/T)}{n^2}.
    \end{eqnarray}
 Finally, we obtain
 \begin{eqnarray}
    &&V_I=V_I^{\text{vacuum}}+V_I^{T} \nonumber \\
    &&-\frac{m^4}{64\pi^2} \biggl(
\ln\left(\frac{\widetilde{\mu}^2}{m^2}\right)+\frac{3}{2} \biggr) -\frac{m^2 T^2}{2\pi^2} \sum_{n=1}^\infty \frac{K_2(nm/T)}{n^2}.\nonumber \\
 \end{eqnarray}
 Now we proceed with the calculation of the term $V_{II}$. First we integrate in  $dm^2$
 \begin{eqnarray}
    V_{II} &\equiv& \frac{T}{2}\sum_n \int dm^2 \int_{-\infty}^{\infty} \frac{d^3p}{(2\pi)^3}(qE)^2 \nonumber \\ 
    &\times&\left(\frac{2(\omega_n^2+p_3^2) }{(\omega_n^2+p^2+m^2)^4} -\frac{1}{(\omega_n^2+p^2+m^2)^3}\right).  \nonumber\\
    &=&\frac{(qE)^2T}{2}\sum_n  \int_{-\infty}^{\infty} \frac{d^3p}{(2\pi)^3} \nonumber \\
    &\times&
    \left(\frac{-2(\omega_n^2+p_3^2) }{3(\omega_n^2+p^2+m^2)^3} +\frac{1}{2(\omega_n^2+p^2+m^2)^2}\right) \nonumber \\
    &=&\frac{(qE)^2T}{2}\sum_n  \int_{-\infty}^{\infty} \frac{d^3p}{(2\pi)^3} \nonumber \\
    &\times&
    \left(-\frac{1}{6(\omega_n^2+p^2+m^2)^2}+\frac{2(p_{\perp}^2+m^2) }{3(\omega_n^2+p^2+m^2)^3} \right) \nonumber \\
    &=&\frac{(qE)^2T}{2}\sum_n  \int_{-\infty}^{\infty} \frac{d^3p}{(2\pi)^3} \Biggl(\frac{4 p^2}{9(\omega_n^2+p^2+m^2)^3} \nonumber \\
    &-& \frac{1}{3}\left(m^2 \frac{\partial}{\partial m^2} +\frac{1}{2} \right)\frac{1}{(\omega_n^2+p^2+m^2)^2} \Biggr), \label{a2}
    \end{eqnarray}
    where we have used
\begin{equation}
 -\frac{1}{2}\frac{\partial}{\partial m^2} \frac{1}{(\omega_n^2+p^2+m^2)^2} = \frac{1}{(\omega_n^2+p^2+m^2)^3},   
\end{equation}
and the fact that $p_{\perp}^2=2p^2/3$ because of spherical symmetry. In this way we get

\begin{equation}
  \frac{1}{2 p^2}\frac{\partial^2}{\partial t^2} \frac{p^2}{(\omega_n^2+tp^2+m^2)}\Biggr|_{t=1} = \frac{p^2}{(\omega_n^2+p^2+m^2)^3} ,\label{parcial1} 
\end{equation}
\begin{equation}
  -\frac{1}{p^2}\frac{\partial}{\partial t} \frac{p^2}{(\omega_n^2+tp^2+m^2)}\Biggr|_{t=1} = \frac{1}{(\omega_n^2+p^2+m^2)^2} , \label{parcial2}
  \end{equation}
  Using the above  Eq.(\ref{parcial1}) and Eq.(\ref{parcial2}) in Eq.(\ref{a2}), we get
        \begin{eqnarray}
    &&V_{II}=\frac{(qE)^2T}{2}\sum_n  \int_{-\infty}^{\infty} \frac{d^3p}{(2\pi)^3} \Biggl[\frac{2\partial^2}{9p^2\partial t^2} \nonumber \\
    &+&
    \frac{\partial}{p^2\partial t}\frac{1}{3}\left(m^2 \frac{\partial}{\partial m^2} +\frac{1}{2} \right)\Biggr]\frac{1}{(\omega_n^2+tp^2+m^2)}\Biggr|_{t=1}.  
\end{eqnarray}
Now we carry out the Matsubara sum, getting
 \begin{eqnarray}
    &&V_{II}=\frac{(qE)^2}{2}   \nonumber \\
    &\times&
    \left[\frac{2\partial^2}{9\partial t^2}+\frac{\partial}{\partial t}\frac{1}{3}\left(m^2 \frac{\partial}{\partial m^2} +\frac{1}{2} \right)\right]\nonumber \\
    &&\int_0^{\infty} 4\pi \frac{dp}{(2\pi)^3}  \frac{1}{2 \sqrt{t p^2+m^2}} (1+2n_B(\sqrt{t p^2+m^2}) \Biggr|_{t=1} . \nonumber \\ 
\end{eqnarray}
If we introduce the change of variable $p=x/\sqrt{t}$,  $dp=dx/\sqrt{t}$, we obtain
 \begin{eqnarray}
    &&V_{II}=\frac{(qE)^2}{2}   \nonumber \\
    &\times&
    \left[\frac{2\partial^2}{9\partial t^2}+\frac{\partial}{\partial t}\frac{1}{3}\left(m^2 \frac{\partial}{\partial m^2} +\frac{1}{2} \right)\right]\nonumber \\
    &&\int_0^{\infty}  \frac{dx}{(2\pi)^2}  \frac{1}{\sqrt{t} \sqrt{x^2+m^2}} (1+2n_B(\sqrt{x^2+m^2}) \Biggr|_{t=1} , 
\end{eqnarray}
 By taking a derivative in  $t$ evaluating then in $t=1$, we obtain
 \begin{eqnarray}
    &&V_{II}=\frac{(qE)^2}{2}   
       \left[\frac{1}{6}-\frac{1}{6}\left(m^2 \frac{\partial}{\partial m^2} +\frac{1}{2} \right)\right]\nonumber \\
    &\times&\int_0^{\infty} \frac{dx}{(2\pi)^2}  \frac{1}{ \sqrt{x^2+m^2}} (1+2n_B(\sqrt{x^2+m^2}) \nonumber \\ 
    &=&\frac{(qE)^2}{12}       \left[\frac{1}{2}-m^2 \frac{\partial}{\partial m^2} \right]\nonumber \\
    &\times&\int_0^{\infty} \frac{dx}{(2\pi)^2}  \frac{1}{ \sqrt{x^2+m^2}} (1+2n_B(\sqrt{x^2+m^2})) \nonumber \\ 
   &=&\frac{(qE)^2}{12}       \left[\frac{1}{2}-m^2 \frac{\partial}{\partial m^2} \right] \int_0^{\infty} \frac{dx}{(2\pi)^2}  \frac{1}{ \sqrt{x^2+m^2}}\nonumber \\
    &+&\frac{(qE)^2}{6 (2\pi)^2}   \left[\frac{1}{2}-m^2 \frac{\partial}{\partial m^2} \right]\sum_{n=0}^{\infty}K_0((n+1)m/T), \nonumber\\
\end{eqnarray}
where in the last equality we have used  Eq. (\ref{bessel}). Taking the mass derivatives we have
 \begin{eqnarray}
    &&V_{II}=\frac{(qE)^2}{12}       \Biggl[\frac{1}{2}  \int_0^{\infty} \frac{dx}{(2\pi)^2}  \frac{1}{ \sqrt{x^2+m^2}}\nonumber \\
   &+&\int_0^{\infty}\frac{dx}{2(2\pi)^2}  \frac{m^2}{ (x^2+m^2)^{3/2}} \Biggr] \nonumber \\
    &+&\frac{(qE)^2}{6 (2\pi)^2}   \Biggl[\sum_{n=1}^{\infty}\frac{K_0(nm/T)}{2}+\frac{m}{2T}\sum_{n=1}^{\infty}nK_1(nm/T) \Biggr].\nonumber \\
  \end{eqnarray}
For the first term in the above expression we use again dimensional regularization. The second term can be integrated finding.
 \begin{eqnarray}
    &&V_{II}=\frac{(qE)^2}{96 \pi^2}       \Biggl[ \frac{2}{\epsilon}
+\ln\left(\frac{\widetilde{\mu}^2}{m^2}\right)  +1 \nonumber \\
   &+& \sum_{n=1}^{\infty} 2K_0(nm/T) + \frac{2m}{T} \sum_{n=1}^{\infty}nK_1(nm/T)  \Biggr].
\end{eqnarray}
Finally, summing the contributions $V_I$ and  $V_{II}$, we obtain the effective potential in the weak electric field region for the whole temperature range
\begin{eqnarray}
&&V=-\frac{m^4}{64\pi^2}  \Biggl(
\ln\left(\frac{\widetilde{\mu}^2}{m^2}\right)+\frac{3}{2} \Biggr)  \nonumber \\
\nonumber \\
&-&\frac{m^2 T^2}{2\pi^2} \sum_{n=1}^\infty \frac{K_2(nm/T)}{n^2} \nonumber \\
&+&\frac{(qE)^2}{96 \pi^2}       \Biggl[ \frac{2}{\epsilon}
+\ln\left(\frac{\widetilde{\mu}^2}{m^2}\right) +1 \nonumber \\
   &+& \sum_{n=1}^{\infty} 2K_0(nm/T) + \frac{2m}{T} \sum_{n=1}^{\infty}nK_1(nm/T)  \Biggr].
\end{eqnarray}

\section{Numerical comparison}\label{apendiceb}
We would like to stress that our expression for the effective potential in the weak electric field intensity region is valid for any value of temperature. This is a novel result. Normally two different analyses were introduced. See \cite{zamora5}. A low temperature discussion as well as a high temperature expansion  were separately introduced and there was  no closed analytic expression for the intermediate temperature region. Here we have  found a single closed analytic expression  valid for any value of temperature. 

\noindent
This point has been checked numerically. As an  example, let us consider the following integral, which is part of the effective potential in the low electric field intensity region and which appears in the appendix A in Eq.~(\ref{A7})

\begin{align}
h(\beta)	&\equiv\int_0^\infty \frac{p^2\;dp}{\sqrt{p^2+m^2}}\frac{1}{e^{\sqrt{p^2+m^2}\beta}-1},
\end{align}
where $\beta=1/T$. Using our methods, this integral can be expressed as

\begin{equation}
    h(\beta)  =\sum_{n=1}^\infty \frac{mT}{n}K_1\left(\frac{mn}{T}\right).
\end{equation}

\begin{figure}[h]
\includegraphics[width=85mm]{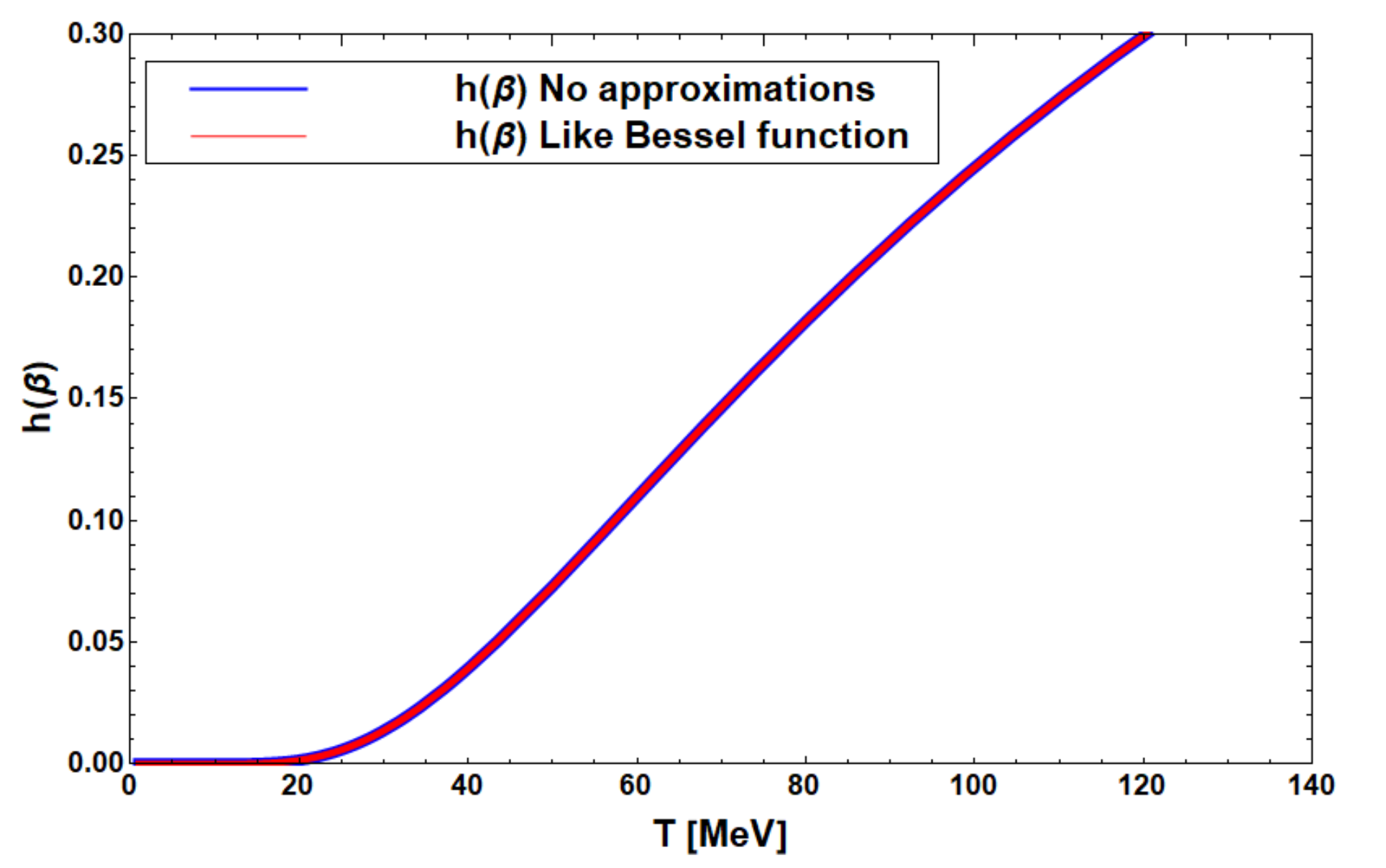} 
\caption{Comparison of the function $h(\beta)$ with the $h(\beta)$ like Bessel function. The blue line corresponds to the numerical evaluation of $h(\beta)$ and the red line to the case of $h(\beta)$ expressed in term of Bessel functions.}
\label{comparacion}
\end{figure}

In Fig.~\ref{comparacion}  we show the curves corresponding to the numerical result and to the analytic expression in terms of a series of Bessel functions. They coincide exactly for any possible temperature value. 

In \cite{zamora5} we had analyzed the same integral, this time in the context of the discussion of thermo-magnetic renormalons and, since we did not have at this time the expansion in terms of Bessel functions, it was necessary to explore separately the low and high temperature regions. This is now no longer needed.

\end{appendix}




\newpage
\begin{thebibliography}{89}

\bibitem{general reviews}. V. A. Miransky and I. A. Shovkovy, Phys. Rept. 576, 1 (2015);  G. Baym.T. Hatsuda, T. Kojo, P.D. Powell, Y. Song, and T. Takasuka, Rept. Prog. Phys. 81 (2018) 5, 056902; D. Blaschke, A. Ayriyan, and A. Friesen (Editors), ``Compact Stars in the QCD Phase Diagram", Universe, MppiaG (2020).  


\bibitem{effective lagrangians} Alejandro Ayala, Luis. A. Hern\'andez, Marcelo Loewe, and Cristian Villavicencio, Eur. Phys. J.A 57 (2021), 7, 234. 
\bibitem{bali01} G. S. Bali, B. B. Brandt, G. Endr{\H o}di and B. Gl{\"a}ssle, Phys. Rev. Lett. {\bf 121}, 072001 (2018).
%
\bibitem{iranianos} Sh. Fayazbakhsh and N. Sadooghi, Phys. Rev. D {\bf 88}, 065030 (2013).

\bibitem{zamora1} A. Ayala, J.L Hern\'andez, L. A. Hern\'andez, R.L.S Farias and R. Zamora, Phys. Rev. D 103, 054038 (2021).

\bibitem{zamora2} A. Ayala, J.L. Hern\'andez, L. A. Hern\'andez, R.L.S Farias and R. Zamora, Phys. Rev. D 102, 114038 (2020).

\bibitem{zamora3} C.A. Dominguez, L. A. Hern\'andez, M. Loewe, C. Villavicencio and   R. Zamora, Phys. Rev. D 102, 094007 (2020).
%
\bibitem{simonov03} Yu. A. Simonov, Phys. At. Nucl. {\bf 79}, 455 (2016).
%
\bibitem{aguirre02} R.~M.~Aguirre,
Eur. Phys. J. A \textbf{55}, 28 (2019).
%
\bibitem{tetsuya} T. Yoshida and K. Suzuki,
Phys. Rev. D {\bf 94}, 074043 (2016).
%
\bibitem{dudal04} D. Dudal and T. G. Mertens,
Phys. Rev. D {\bf 91}, 086002 (2015).
%
\bibitem{kevin} K.~Marasinghe and K.~Tuchin,
Phys. Rev. C {\bf 84}, 044908 (2011).
%
\bibitem{gubler} P. Gubler, K. Hattori, S. H.ff Lee, M. Oka, S. Ozaki and K. Suzuki, Phys. Rev. D {\bf 93}, 054026 (2016).
%
\bibitem{noronha01} C. S. Machado, S. I. Finazzo, R. D. Matheus and J. Noronha, Phys. Rev. D {\bf 89}, 074027 (2014).
%
\bibitem{morita} S. Cho, K. Hattori, S. H. Lee, K. Morita and S. Ozaki,
Phys. Rev. Lett. {\bf 113}, 172301 (2014).


\bibitem{Ayala1} A. Ayala, R. L. S. Farias, S. Hern\'andez-Ort\'iz, L. A. Hern\'andez, D. Manreza Paret and R. Zamora, Phys. Rev. D {\bf 98}, 114008 (2018).
%
\bibitem{morita02}  S. Cho, K. Hattori, S. H. Lee, K. Morita and S. Ozaki,
Phys. Rev. D {\bf 91}, 045025 (2015).
%
\bibitem{sarkar03} S.~Ghosh, A.~Mukherjee, M.~Mandal, S.~Sarkar and P.~Roy,
Phys. Rev. D \textbf{94}, 094043 (2016).
%
\bibitem{band} A.~Bandyopadhyay and S.~Mallik,
Eur. Phys. J. C \textbf{77}, 771 (2017).
%

%
\bibitem{nosso1} S. S. Avancini, W. R. Tavares and M. B. Pinto, Phys. Rev. D {\bf 93}, 014010 (2016).
%
\bibitem{nosso03} S. S. Avancini, R. L. Farias, M. B. Pinto, W. R. Tavares and V. S. Timteo, Phys. Lett. B {\bf 767}, 247 (2017).



\bibitem{Ayala2} A. Ayala, C. A. Dominguez, L. A.
Hern\'andez, M. Loewe and R. Zamora, Phys. Rev. D {\bf 92}, 096011 (2015).


\bibitem{zamora4} A. Ayala, S. Hernandez-Ortiz, L. A. Hernandez, V. Knapp-Perez and R. Zamora, Phys. Rev. D 201, 074023 (2020).

\bibitem{zamora5} M. Loewe, L. Monje and R. Zamora, Phys. Rev. D 104, 016020 (2021).

\bibitem{peng}
M. Ruggieri and G. X. Peng, Phys. Rev. D 93, 094021 (2016).

\bibitem{cohen}
D. Cohen, D. A. McGady, and E. S. Werbos, Phys. Rev. C 76, 055201 (2007).

\bibitem{tavares}
W. R. Tavares and S. S. Avancini, Phys. Rev. D 97, 094001 (2018).



\bibitem{25}
W. T. Deng and X. G. Huang, Phys. Lett. B 742, 296 (2015).

\bibitem{26}
Y. Hirono, M. Hongo, and T. Hirano, Phys. Rev. C 90, 021903(R) (2014).

\bibitem{27}

V. Voronyuk, V. D. Toneev, S. A. Voloshin, and W. Cassing, Phys. Rev. C 90, 064903 (2014).





\bibitem{Dittrich} Walter Dittrich and Martin Reuter, ``Efective Lagrangiand in Quantum Electrodynamics". Springer-Verlag, Berlin Heidelberg New York, Tokyo (1985). See also Walter Dittrich and Holger Gies, ``Probing the Quantum Vacuum: Perturbative effective action approach in Queantum Electrodynamics and its application", Springer Tracts in Modern Physics, Volume 166 (2000)

\bibitem{ahmad}
A. Ahmad, N. Ahmadiniaz, O. Corradini, S. P. Kim and C. Schubert, Nuclear Physics B, 919 (2017).


\bibitem{ayalaeuro}
A. Ayala, J. Casta\~no, L. A. Hernandez, J. Salinas and R. Zamora, Eur. Phys. J. A57, (2021).

\bibitem{taiwaneses}
T.-K. Chyi, C.-W. Hwang, W. F. Kao, G.-L. Lin, and K.-W. Ng, Phys. Rev. D {\bf 62}, 105014 (2000).

\bibitem{mexicanos}
A. Ayala, A. Sanches, G. Piccinelli and S. Sahu, Phys. Rev. D 71, 023004 (2005).

\bibitem{Nambu} Y. Nambu, Phys. Rev. {\bf 117} 648 (1960).

\bibitem{abramowitz}
Abramowitz, M. and Stegun, I. A. (Eds.). Handbook of Mathematical Functions with Formulas, Graphs, and Mathematical Tables, 9th printing. New York: Dover, pp. 576-579, 1972.

\bibitem{LeBellac}
M. Le Bellac, {\it Thermal Field Theory}, Cambridge University Press,
1996.

\bibitem{Kapusta}
J. I. Kapusta and C. Gale, {\it Finite-Temperature Field Theory Principles and Applications}, Cambridge University Press,
2006.

\bibitem{inverse1}
A. Ayala, M. Loewe and R. Zamora, Phys. Rev. D91, 016002 (2015).

\bibitem{inverse2}
A. Ayala, M. Loewe, A. Mizher and R. Zamora, Phys. Rev. D90, 036001 (2014).

\bibitem{inverse3}
Pedro Costa, Márcio Ferreira, Débora P. Menezes, Jo\~ao
Moreira, and Constança Provid\^encia, Phys. Rev. D,
92, 036012 (2015).

\bibitem{inverse4}
Jens O. Andersen, Eur. Phys. J. A (2021) 57: 189.

\bibitem{inverse5}
A. Ahmad and A. Raya, J. Phys. G43, 065002, 2016.

\bibitem{inverse6}
M. Ferreira, P. Costa, O. Lourenç0, T. Fredetico and C. Provid\^encia, Phys. Rev. D,
89, 116011 (2014).

\bibitem{inverse7}
A. Ayala, L. A. Hernández, M. Loewe, J. C. Rojas and R. Zamora, Eur. Phys. J. A56, (2020).

\bibitem{Farias} W. R. Tavares, R. L. S. Farias and S. S. Avancini, Phys. Rev. D {\bf 101}, 016017 (2020).










 







\end{thebibliography}
\end{document}